\documentstyle[12pt]{article}

\newcommand {\vs}[1]  { \vspace*{#1 cm} }

\newcounter{eq}
\newcounter{sc}

\newcommand {\MPL}  {Mod. Phys. Lett.}
\newcommand {\IJMP}  {Int. J. Mod. Phys.}
\newcommand {\NP}   {Nucl. Phys.}

\newcommand {\PL}   {Phys. Lett.}
\newcommand {\PRP}  {Phys. Rept.}
\newcommand {\PR}   {Phys. Rev.}

\newcommand {\AP}   {Ann. of Phys.}

\def\overleftrightarrow#1{\vbox{\ialign{##\crcr
 $\leftrightarrow$\crcr\noalign{\kern-1pt\nointerlineskip}
 $\hfil\displaystyle{#1}\hfil$\crcr}}}

\newlength{\minitwocolumn}
\begin{document}

\begin{flushright}
DPUR/TH/11\\
August, 2008\\
\end{flushright}
\vspace{30pt}
\pagestyle{empty}
\baselineskip15pt

\begin{center}
{\large\bf Massive Gravity in Curved Cosmological Backgrounds

 \vskip 1mm
}

\vspace{20mm}

Masahiro Maeno
          \footnote{
           E-mail address:\ maeno@sci.u-ryukyu.ac.jp
                  }
and Ichiro Oda
          \footnote{
           E-mail address:\ ioda@phys.u-ryukyu.ac.jp
                  }

\vspace{10mm}
          Department of Physics, Faculty of Science, University of the 
           Ryukyus,\\
           Nishihara, Okinawa 903-0213, JAPAN \\

\end{center}


\vspace{20mm}
\begin{abstract}
We study the physical propagating modes in a massive gravity model in curved
cosmological backgrounds, which we have found as classical solutions in our 
previous paper. We show that, generically, there exist such the cosmological 
background solutions consistent with the equations of motion where we assume 
the ghost condensation ansatzes. Using the (1+3)-parametrization of the metric 
fluctuations with 'unitary' gauge, we find that there is neither a scalar 
ghost nor a tachyon in the spectrum of the propagating modes, the tensor 
modes become massive owing to gravitational Higgs mechanism, and the model 
is free of the Boulware-Deser instability. The price we have to pay is that 
the scalar sector breaks the Lorentz-invariance, but there are no pathologies 
in the spectrum and lead to interesting phenomenology. Moreover, we present 
a proof of the absence of non-unitary modes for a specific ghost condensation 
model in a cosmological background.

\vspace{15mm}

\end{abstract}

\newpage
\pagestyle{plain}
\pagenumbering{arabic}


\rm
\section{Introduction}

In recent years, we have watched revival of interests to construct massive
gravity theories from different physical motivations \cite{Percacci}-\cite{Rubakov2}. 
It concerns a very simple question: Can the graviton have a (small) mass consistent with
experiments?  If so, how?

One motivation behind this question comes from the astonishing 
observational fact that our universe is not just expanding but is at present 
in an epoch of undergoing an accelerating expansion \cite{Riess, Perlmutter}. 
Although the standard model of cosmology is remarkably successful 
in accounting for many of observational facts of the universe, 
it is fair to say that we are still 
lacking a fundamental understanding of the late-time cosmic acceleration 
in addition to problems associated with dark matter and dark energy. 

Several ideas have been thus far put forward for explaining this 
perplexing fact \cite{Copeland}.  It is here that massive gravity theories might play 
an important role since massive gravity theories could modify Einstein's theory 
of general relativity at large cosmological scales and might lead to 
the present accelerated expansion of the universe without introducing 
still mysterious dark matter and dark energy at all. Since general relativity
is almost the unique theory of massless spin 2 gravitational field
whose universality class is determined by local symmetries under general
coordinate transformations, any infrared modification of general relativity
cannot aviod introduction of some kind of mass for the graviton.

The other motivation for attempting to construct massive gravity theories
lies in string theory approach to quantum chromodynamics (QCD) \cite{'t Hooft}.
For instance, as inspired in the large-N expansion of the gauge theory,
which defines the planar (or genus zero) diagram and is analogous to the tree
diagram of string theory, if we wish to apply a bosonic string theory 
to the gluonic sector in QCD, massless fields such as spin 2 graviton
in string theory, must become massive or be removed somehow by an ingenious 
dynamical mechanism since such the massless fields do not appear in QCD.
Note that this motivation is relevant to the modification of general relativity
in the short distance region while the previous cosmological one is to the 
infrared modification. It is worthwhile to point out that the modification of 
both short and long distance regions reminds us of T-duality $R \leftrightarrow l_s^2/R$ 
in string theory where $l_s$ is the fundamental string length scale.
If there is the interesting symmetry behind the both distance regions
in general relativity, such a modification could perhaps open a fruitful dialog 
between cosmology and QCD.

An approach for the construction of massive gravity theories is to take account 
of the spontaneous symmetry breakdown (SSB) of general coordinate 
reparametrization invariance \cite{Percacci}-\cite{'t Hooft}.
Note that the SSB of general coordinate reparametrization invariance might provide 
a resolution for cosmological constant problem \cite{Kirsch, 't Hooft}. 
Though the cosmological constant problem needs a theory of quantum gravity, 
solving the problem requires a low energy mechanism, for instance, a partial 
cancellation of vacuum energy density stemming from quantum loops. 
We expect that from the analogy with the Higgs mechanism in conventional gauge theories, 
the Nambu-Goldstone mode mixes with the massless graviton, thereby changing the vacuum 
structure of gravitational sector in a non-trivial way.    

Recently, as a device for performing a consistent infrared modification of general relativity
an idea of ghost condensation has been proposed \cite{Arkani}. In this scenario, 
the '$\it{unitary}$' propagating scalar field appears as the Nambu-Goldstone boson 
for a spontaneously broken time-like diffeomorphism, 
and yields a possibility for resolving cosmological problems such as inflation, 
dark matter and dark energy. The key observation here is that a scalar ghost is converted 
to a normal scalar with positive-energy excitations.  

More recently, 't Hooft has proposed a new Higgs mechanism for gravity
where the massless graviton '$\it{eats}$' four real scalar fields and consequently
becomes massive \cite{'t Hooft}. In his model, vacuum expectation values (VEV's) 
of the scalar fields are taken to be the four space-time coordinates by gauge-fixing 
diffeomorphisms, so the whole diffeomorphisms are broken spontanously. 
Of course, the number of dynamical degrees of freedom is left unchanged 
before and after the SSB. Actually, before the SSB of diffeomorphisms 
there are massless gravitons of two dynamical degrees of freedom and 
four real scalar fields whereas after the SSB we have massive gravitons of five 
dynamical degrees of freedom and one real scalar field. 
Afterward, a topological term was included to the 't Hooft model 
where an '$\it{alternative}$' metric tensor is naturally derived and the topological meaning 
of the gauge conditions was clarified \cite{Oda}. \footnote{Similar but different approaches 
have been already taken into consideration in Ref. \cite{Oda1}.}  

The problem in the 't Hooft model is that a scalar field appearing after the SSB is a non-unitary 
propagating field so that in order to avoid violation of unitarity
it must be removed from the physical Hibert space in terms of some 
procedure. \footnote{More recently, models of gravitational Higgs mechanism 
without the non-unitary propagating scalar field were proposed
in Ref. \cite{Kaku2}.} 
A resolution for this problem was offered where one requires the energy-momentum tensor of 
the matter field to couple to not the usual metric tensor but the modified metric one 
in such a way that the non-unitary scalar field does not couple to the energy-momentum 
tensor directly \cite{'t Hooft}. 

In this context, one could imagine that the ghost condensation scenario enables us 
to avoid emergence of the non-unitary scalar field in gravitational Higgs mechanism 
since the annoying non-unitary scalar field has its roots in one time-like component 
(i.e., ghost) in four scalar fields, to which the ghost condensation idea could be applied. 
Thus, it should be regarded that this observation is an alternative method for removing 
the non-unitary propagating scalar field in gravitational Higgs mechanism by 't Hooft.

In our previous article \cite{Maeno}, as a first step for proving this conjecture,
we have studied classical solutions in the unitary gauge in general ghost condensation models. 
This analysis is needed for understanding in what background gravitational 
Higgs mechanism arises. It turns out that 
depending on the form of scalar fields in an action, there are three kinds of classical, 
exact solutions, which are (anti-) de Sitter space-time, polynomially expanding universes and flat 
Minkowski space-time. 

In this article, we wish to prove that there is indeed no non-unitary mode
in the spectrum of propagating modes around cosmological backgrounds
and the graviton becomes massive because of gravitational Higgs mechanism in these models.
Thus, the models at hand are free of the problem associated with the non-unitary propagating 
mode and the Boulware-Deser instability \cite{Boulware} in curved backgrounds.

This paper is organized as follows: In section 2, we review that there is an
expanding cosmological solution with zero acceleration to the equations 
of motion in a massive gravity model, what we call, 't Hooft model with ghost condensation. 
In section 3, we consider a more general model and look for classical solutions where we will 
find expanding cosmological solutions with non-zero acceleration. In section 4, based on 
the massive gravity model in section 3, the propagating modes around the cosmological
backgrounds are examined in detail by using the (1+3)-parametrization of the metric 
fluctuations. We find that the tensor modes become massive, 
the vector modes are non-dynamical, three of the scalar modes are also non-dynamical
and one scalar becomes massive. This scalar is originally non-unitary mode,
that is, a ghost, but becomes a normal particle because of the ghost
condensation mechanism. However, the dispersion relation is not usual
one, so this mode breaks the Lorentz invariance as expected from
the ghost condensation scenario. In section 5, we present a proof of the absence 
of non-unitary modes for the specific ghost condensation model treated in section 2.
The final section is devoted to conclusions and discussion.

\section{Cosmological solution in 't Hooft model with ghost condensation}

In this section, we wish to review our previous work \cite{Maeno} 
and explain how to derive classical solutions in massive gravity models.

Let us start with review of our 't Hooft model with ghost condensation.
The general proof that there is no non-unitary scalar ghost in this
model will be given in section 5.
The action of ghost condensation \cite{Arkani}, which is inspired by 
gravitational Higgs mechansim, takes the form \footnote{For simplicity, we have 
put the Newton constant $G_N$ at the front, so the dimension of ghost-like scalar field 
$\phi^0$ differs from that of
the original ghost condensation model, but it is easy to modify the dimension by
field redefinition.}:
\begin{eqnarray}
S &=& \frac{1}{16 \pi G_N} \int d^4 x \sqrt{-g} [ R - 2 \Lambda + f(X) 
- g^{\mu\nu} \partial_\mu \phi^a \partial_\nu \phi^b \eta_{ab} ]  
\nonumber\\
&=& \frac{1}{16 \pi G_N} \int d^4 x \sqrt{-g} [ R - 2 \Lambda + F(X) 
- g^{\mu\nu} \partial_\mu \phi^i \partial_\nu \phi^j \delta_{ij} ],
\label{Main}
\end{eqnarray}
where $\eta_{ab}$ is the internal flat metric with diagonal elements $(-1, +1, +1, +1)$. 
The indices $\mu, \nu (= 0, 1, 2, 3)$ and $i, j (= 1, 2, 3)$ are space-time 
and spatial indices, respectively.
And we have defined $F(X) \equiv X + f(X)$ where $f(X)$ is a function of 
$X$ which is defined as
\begin{eqnarray}
X = g^{\mu\nu} \partial_\mu \phi^0 \partial_\nu \phi^0.
\label{X}
\end{eqnarray}
Let us note that the last term in the first equality in (\ref{Main}) is added 
to the original ghost condensation action, 
or alternatively speaking from the side of gravitational Higgs mechanism, 
the third term $f(X)$ is introduced to the original action of gravitational 
Higgs mechanism.

Recall that the ghost condensation scenario consists of three ansatzes:
The first ansatz amounts to requiring that the function $F(X)$ has 
a minimum at some point $X_{min}$ such that
\begin{eqnarray}
F_X(X_{min}) = 0, \  F_{XX}(X_{min}) > 0,
\label{Mini}
\end{eqnarray}
where $F_X(X_{min})$, for instance, means the differentiation of $F(X)$ 
with respect to $X$ and then putting $X = X_{min}$. Note that the latter 
condition in (\ref{Mini}) ensures ghost condensation.

As the second ansatz, the ghost-like scalar field $\phi^0$ is expanded
around the background $m t$ as
\begin{eqnarray}
\phi^0 = m t + \pi,
\label{Pi}
\end{eqnarray}
where $\pi$ is the small fluctuation. This equation can be interpreted
as follows: Using a time-like diffeomorphism $\delta \phi^0 = \varepsilon^0$, 
one can take the gauge $\pi =0$. In other words, $\pi$ is a Nambu-Goldstone 
boson associated with spontaneous symmetry breakdown of time traslation. 

The last ansatz is natural from the second ansatz. That is, since 
time coodinate $t$ plays a distinct role from spatial coordinates
$x^i(i=1, 2, 3)$, it is plausible to make an assumption on background
space-time metric, which is of the Friedmann-Robertson-Walker form
\begin{eqnarray}
ds^2 = - dt^2 + a(t)^2 d \Omega^2,
\label{FRW}
\end{eqnarray}
where $a(t)$ is the scale factor and $d \Omega^2$ is the spatial metric
for a maximally symmetric three-dimensional space.

The equations of motion are easily derived as follows:
\begin{eqnarray}
\partial_\mu (\sqrt{-g} g^{\mu\nu} \partial_\nu \phi^i) &=& 0,
\nonumber\\
\partial_\mu (\sqrt{-g} g^{\mu\nu} F_X(X) \partial_\nu \phi^0) &=& 0,
\nonumber\\
G_{\mu\nu} + \Lambda g_{\mu\nu} &=& T_{\mu\nu},
\label{Eqs1}
\end{eqnarray}
where the stress-energy tensor is defined as
\begin{eqnarray}
T_{\mu\nu} &=&  ( \partial_\mu \phi^a \partial_\nu \phi^b 
- \frac{1}{2} g_{\mu\nu} g^{\alpha\beta} \partial_\alpha \phi^a
\partial_\beta \phi^b ) \eta_{ab}  \nonumber\\
&-& ( \partial_\mu \phi^0 \partial_\nu \phi^0 f_X(X)
- \frac{1}{2} g_{\mu\nu} f(X) ).
\label{Stress1}
\end{eqnarray}

Now let us solve these equations of motion under the ansatzes of ghost condensation 
and the $\it{unitary}$ gauge for diffeomorphisms
\begin{eqnarray}
\phi^a = m x^\mu \delta_\mu^a.
\label{Gauge}
\end{eqnarray}
It is easy to check that the $\phi^i$-equations are trivially satisfied. 
Moreover, the $\phi^0$-equation reduces to the expression
\begin{eqnarray}
\partial_0 ( a(t)^3 F_X(X) ) = 0,
\label{phi0}
\end{eqnarray}
whose validity requires us that one of ghost condensation ansatzes
should be automatically satisfied, i.e., $F_X(X_{min}) =0$ since
$X_{min} \equiv X(\phi^0 = m t) = - m^2$ and $\partial_0 X_{min} = 0$.

Finally, Einstein's equations are cast to 
\begin{eqnarray}
3 (\frac{\dot a}{a})^2 - \tilde \Lambda &=& \frac{3 m^2}{2 a^2},
\nonumber\\
- 2 \frac{\ddot a}{a} - (\frac{\dot a}{a})^2 + \tilde \Lambda 
&=& - \frac{m^2}{2 a^2},
\label{Reduced Eins}
\end{eqnarray}
where we have defined $\tilde \Lambda = \Lambda - \frac{1}{2} F(-m^2)$
and the dot over $a(t)$ denotes the derivative with respect to $t$.
Here we have made use of the expression of the stress-energy tensor 
which is obtained by inserting Eq. (\ref{Gauge}) to Eq. (\ref{Stress1})
\begin{eqnarray}
T_{\mu\nu} &=&  m^2 ( \eta_{\mu\nu} - \frac{1}{2} g_{\mu\nu} g^{ab} \eta_{ab} )
+ m^2 \delta_\mu^0 \delta_\nu^0 + \frac{1}{2} g_{\mu\nu} f(-m^2).
\label{Modified Stress2}
\end{eqnarray}
{}From the equations (\ref{Reduced Eins}), we can eliminate the terms involving $\dot a$ 
whose result is written as
\begin{eqnarray}
3 \frac{\ddot a}{a} = \tilde \Lambda.
\label{Reduced Eins2}
\end{eqnarray}

It then turns out that the general solution for this equation exists only
when $\tilde \Lambda = 0$. Thus, Eq. (\ref{Reduced Eins2}) reads 
$\ddot a = 0$, so we have the general solution
\begin{eqnarray}
a(t) = c (t - t_0),
\label{Solution2}
\end{eqnarray}
where $c$ and $t_0$ are integration constants. Substituting this solution into
the first equation in Eq. (\ref{Reduced Eins}), $c$ is fixed to 
$c = \pm \frac{m}{\sqrt{2}}$. As a result, the line element takes the form
\begin{eqnarray}
d s^2 = - d t^2 + \frac{m^2}{2} (t - t_0)^2 d \Omega^2,
\label{FRW2}
\end{eqnarray}
which describes the linearly expanding universe with zero acceleration.
\footnote{This solution was also obtained in a different massive gravity
model \cite{Kaku2}.}
Notice that this solution is a unique classical solution, so a flat
Minkowski space-time, for instance, is not a solution of this model.
Here it is worthwhile to mention that although the present universe seems to
be accelerating so that the linearly expanding universe is excluded from
WMAP experiment \cite{WMAP}, the general solution (\ref{FRW2}) might be useful in describing 
the future or past status of universe.

\section{Cosmological solutions in more general massive gravity models}

In the previous section, we have found an interesting cosmological solution,
which is very similar to the Milne universe, in the 't Hooft model with ghost 
condensation. However, the solution has zero acceleration so that it
does not describe the present epoch of the accelerating universe.
It would be more desirable if we could get a classical solution with
non-zero acceleration. It has been shown in our previous paper \cite{Maeno}
that this desire is actually realized if one generalizes the model in section 2
to, what is called, a general ghost condensation model. In this
article, instead of explaining the derivation of the solutions in
the general ghost condensation model, we show that the solutions 
also exist in more general massive gravity models with the ghost condensation 
potential. 

In this section, we will obey the following line of arguments:
In order to make the analysis of propagating modes easier, which will be done
in the next section, we will first choose a simple massive gravity model 
in a general class of those models. Next, using this specific model, we will
show that there are polynomially expanding universes' solutions with
non-zero acceleration. Finally, we will examine under what conditions 
the solutions at hand remain the solutions to the equations of motion
in the most general models.

Now we shall take a different line element from (\ref{FRW})
since the form is more convenient for later analysis in section 4.
Of course, at a final stage it is easy to transform to the expression
like (\ref{FRW}) through a global coordinate transformation.
The form we take as the line element is of conformal flat type:
\begin{eqnarray}
ds^2 &=& g_{\mu\nu} d x^\mu d x^\nu   
\nonumber\\
&=& e^{2 A(t)} \eta_{\mu\nu} d x^\mu d x^\nu,
\label{C-metric}
\end{eqnarray}
where $\eta_{\mu\nu}$ is the flat metric. 

We consider a simple massive gravity model \footnote{For the analogy with the
previous model in section 2, we have introduced two different functions
$F(X)$ and $G(W^{ij})$ separately for variables $X$ and $W^{ij}$, but
it is more economical to do only one function $F(X, W^{ij})$ dependent on 
two variables. This simple model has been investigated in Refs.
\cite{Dubovsky1}-\cite{Rubakov2}.}:
\begin{eqnarray}
S &=& \frac{1}{16 \pi G_N} \int d^4 x \sqrt{-g} [ R - 2 \Lambda + F(X) 
- G(W^{ij}) ]  
\nonumber\\
&\equiv& \frac{1}{16 \pi G_N} \int d^4 x [ L_{EH} + L_\Lambda + L_F
+ L_G ],
\label{Main2}
\end{eqnarray}
where $X$ is defined in (\ref{X}) and $W^{ij}$ is defined via new
variables $Y^{ij}$ and $V^i$ as
\begin{eqnarray}
W^{ij} &=& Y^{ij} - \frac{V^i V^j}{X}, \nonumber\\
Y^{ij} &=& g^{\mu\nu} \partial_\mu \phi^i \partial_\nu \phi^j,
\nonumber\\
V^i &=& g^{\mu\nu} \partial_\mu \phi^0 \partial_\nu \phi^i.
\label{W}
\end{eqnarray}
This model of massive gravity was obtained by considering the
residual diffeomorphisms
\begin{eqnarray}
x^i \rightarrow x'^i = x^i + \zeta^i(t),
\label{Res-diff}
\end{eqnarray}
which are translated to the symmetries of the scalar fields as
\begin{eqnarray}
\phi^i \rightarrow \phi'^i = \phi^i + \xi^i(\phi^0),
\label{Res-diff2}
\end{eqnarray}
where $\zeta^i(t)$ and $\xi^i(\phi^0)$ are the infinitesimal transformation
parameters which are dependent on $t$ and $\phi^0$, respectively. Note
that both $X$ and $W^{ij}$ are invariant under (\ref{Res-diff2}).

Let us show that there are polynomially expanding cosmological
solutions with non-zero acceleration in this model. To this end, we first
derive the equations of motion. The $\phi^i$-, $\phi^0$-, and Einstein's
equations are derived in a straightforward manner
\begin{eqnarray}
\partial_\mu [\sqrt{-g} g^{\mu\nu} \frac{\partial G}{\partial W^{ij}} 
(\partial_\nu \phi^j - \frac{V^j}{X} \partial_\nu \phi^0)] &=& 0,
\nonumber\\
\partial_\mu [\sqrt{-g} g^{\mu\nu} \{ \partial_\nu \phi^0 F_X +
(\frac{V^j}{X} \partial_\nu \phi^i - \frac{V^i V^j}{X^2} \partial_\nu \phi^0)
\frac{\partial G}{\partial W^{ij}} \}] &=& 0,
\nonumber\\
G_{\mu\nu} + \Lambda g_{\mu\nu} &=& T_{\mu\nu},
\label{Whole Eq}
\end{eqnarray}
with the stress-energy tensor being given by
\begin{eqnarray}
T_{\mu\nu} &=& - \partial_\mu \phi^0 \partial_\nu \phi^0 F_X 
+ [ \partial_\mu \phi^i \partial_\nu \phi^j
- \frac{2}{X} \partial_\mu \phi^0 \partial_\nu \phi^i V^j 
+ \frac{V^i V^j}{X^2} \partial_\mu \phi^0 \partial_\nu \phi^0]
\frac{\partial G}{\partial W^{ij}}
\nonumber\\
&+& \frac{1}{2} g_{\mu\nu} [ F - G ].
\label{Stress2}
\end{eqnarray}

With the $\it{unitary}$ gauge (\ref{Gauge}), each variable takes the
form
\begin{eqnarray}
X &=& m^2 g^{00} = - m^2 e^{-2A} \equiv \bar{X},
\nonumber\\
V^i &=& m^2 g^{0i} = 0 \equiv \bar{V}^i,
\nonumber\\
Y^{ij} &=& m^2 g^{ij} = m^2 e^{-2A} \delta_{ij} \equiv \bar{Y}^{ij},
\nonumber\\
W^{ij} &=& m^2 e^{-2A} \delta_{ij} \equiv \bar{W} \delta_{ij}
\equiv \bar{W}^{ij}.
\label{Variable}
\end{eqnarray}
Then, the $\phi^i$-equations are reduced to 
\begin{eqnarray}
\partial_i [ e^{2A} \frac{\partial G(\bar{W}^{ij})}{\partial W^{ij}} ] = 0,
\label{generalized phi-i}
\end{eqnarray}
which is trivially satisfied since the quantity in the square bracket
depends on only $t$.  

Next, the $\phi^0$-equation reads
\begin{eqnarray}
\partial_t [ e^{2A} F_X (\bar{X}) ] = 0,
\label{generalized phi0}
\end{eqnarray}
which is satisfied by one of ghost condensation ansatzes, $F_X (\bar{X}) = 0$.
Furthermore, Einstein's equations are recast to 
\begin{eqnarray}
&{}& \ddot{A} ( -2 \delta^0_\mu \delta^0_\nu - 2 \eta_{\mu\nu} )
+ ( \dot{A} )^2 ( 2 \delta^0_\mu \delta^0_\nu - \eta_{\mu\nu} )
+ \Lambda e^{2A} \eta_{\mu\nu}
\nonumber\\
&=& m^2 \delta^i_\mu \delta^j_\nu 
\frac{\partial G(\bar{W}^{ij})}{\partial W^{ij}}
+ \frac{1}{2} e^{2A} \eta_{\mu\nu} [ F(\bar{X}) - G(\bar{W}^{ij}) ].
\label{generalized Eins}
\end{eqnarray}

In order to proceed further, we need to fix the form of the
potential term $G(W^{ij})$. It then turns out that there are non-trivial
cosmological solutions provided that we select
\begin{eqnarray}
G(W^{ij}) = \frac{1}{3} K Tr (W^{ij})^n - 2 \tilde \Lambda,
\label{Ansatz G}
\end{eqnarray}
where $K$ is a positive constant and $\tilde \Lambda \equiv \Lambda - \frac{1}{2} 
F(\bar{X})$. For later convenience, using this expression (\ref{Ansatz G})
let us define the following two quantities
\begin{eqnarray}
\frac{\partial G(\bar{W}^{ij})}{\partial W^{ij}} &=& G_1 \delta_{ij},
\nonumber\\
\frac{\partial^2 G(\bar{W}^{ij})}{\partial W^{ij} \partial W^{kl}} 
&=& G_2 (\delta_{ik} \delta_{jl} + \delta_{il} \delta_{jk}),
\label{G}
\end{eqnarray}
where $G_1$ and $G_2$ are defined as
\begin{eqnarray}
G_1 &=&  \frac{K}{3} n \bar{W}^{n-1} 
= \frac{K}{3} n (m^2 e^{-2A})^{n-1},
\nonumber\\
G_2 &=&  \frac{K}{3} \frac{n(n-1)}{2} \bar{W}^{n-2} 
= \frac{K}{3} \frac{n(n-1)}{2} (m^2 e^{-2A})^{n-2}.
\label{GG}
\end{eqnarray}

As a result, Einstein's equations read
\begin{eqnarray}
(\dot{A})^2 &=& - \frac{1}{6} [ F(\bar{X}) - G(\bar{W}^{ij})
- 2 \Lambda ] e^{2A},
\nonumber\\
\ddot{A} + \frac{1}{2} (\dot{A})^2 &=& - \frac{1}{4} [ F(\bar{X}) 
- G(\bar{W}^{ij}) - 2 \Lambda ] e^{2A} - \frac{1}{2} m^2 G_1.
\label{RE.Ein}
\end{eqnarray}
These equations are easily integrated to be 
\begin{eqnarray}
e^A = [ \pm \sqrt{\frac{K}{6}} (n-1) m^n (t - t_0) ]^{\frac{1}{n - 1}},
\label{generalized sol}
\end{eqnarray}
where $t_0$ is an integration constant. Then, the line element becomes (after
a suitable redefinition of $x^\mu$ by an overall constant factor)
\begin{eqnarray}
ds^2 &=& (t - t_0)^{\frac{2}{n-1}} \eta_{\mu\nu} d x^\mu d x^\nu   
\nonumber\\
&=& - d \tau^2 + (\tau - \tau_0)^{\frac{2}{n}} d x^i d x_i,
\label{L-metric}
\end{eqnarray}
where the latter expression informs us that the second derivative of the scale factor, 
that is, the acceleration, is positive for $n < 1$.  Of course, this condition
might not be meaningful phenomenologically since the present universe seems to be
entering a new era of $\it{exponentially}$ accelerating inflation again,
which is controled by the equation of state $w = \frac{P}{\rho} = -1$,
whereas our solutions describe the $\it{polynomially}$ accelerating universes.
Nevertheless, we shall take account of this condition since at present
there do not exist sufficient evidences to exclude the $\it{polynomially}$ 
accelerating universe such as quintessence ($-1 < w < -\frac{1}{3}$)
and phantom ($w < -1$) etc.

In this way, we have obtained an interesting class of cosmological solutions, 
which have a behavior of polynomially expanding universes with non-zero acceleration, 
by generalizing a massive gravity model in such a way that the potential terms 
in the action include functions of not only $\phi^0$ but also $\phi^i$.

To close this section, it is valuable to ask ourselves whether such
the solutions exist even in the most general models or not. Recall
that the symmetries (\ref{Res-diff2}) constrain the model to some degree
in the sense that the potential terms are not a general function of $X$,
$V^i$ and $Y^{ij}$ but the more restricted function of $X$ and $W^{ij}$ 
which are invariant under (\ref{Res-diff2}). 

We therefore take the potential term to be the most general function,
which is an arbitrary function of $X$, $V^i$ and $Y^{ij}$:
\begin{eqnarray}
S = \frac{1}{16 \pi G_N} \int d^4 x \sqrt{-g} [ R - 2 \Lambda 
+ U(X, V^i, Y^{ij}) ].
\label{Main3}
\end{eqnarray}
Now let us derive the equations of motion under the unitary gauge (\ref{Gauge}) 
whose concrete expressions are given by
\begin{eqnarray}
\partial_\mu [\sqrt{-g} ( \frac{1}{2} g^{\mu0} \frac{\partial U}{\partial V^i} 
+ g^{\mu j} \frac{\partial U}{\partial Y^{ij}} )] &=& 0,
\nonumber\\
\partial_\mu [\sqrt{-g} ( g^{\mu0} \frac{\partial U}{\partial X} 
+  \frac{1}{2} g^{\mu i} \frac{\partial U}{\partial V^i} )] &=& 0,
\nonumber\\
G_{\mu\nu} + \Lambda g_{\mu\nu} &=& T_{\mu\nu}.
\label{Whole Eq2}
\end{eqnarray}
Now, the second $\phi^0$-equation is satisfied if one imposes the
ghost condensation ansatz
\begin{eqnarray}
\frac{\partial U(\bar{X}, \bar{V}^i, \bar{Y}^{ij})}{\partial X} = 0.
\label{New ansatz}
\end{eqnarray}
Next, we find that the first $\phi^i$-equations are satisfied if and
only if there exist new conditions on $U$ 
\begin{eqnarray}
\frac{\partial U(\bar{X}, \bar{V}^i, \bar{Y}^{ij})}{\partial V^i} = 0.
\label{New ansatz2}
\end{eqnarray}
In this regard, note that in order to keep the $SO(3)$ rotational symmetry, 
$U$ must be a function of $V^i V^i$, $V^i Y_{ij} V^j$ etc., 
so $\frac{\partial U(\bar{V}^i)}
{\partial V^i}$ would be proportional to $\bar{V}^i$. In the unitary gauge,
$\bar{V}^i$ is vanishing so that the conditions (\ref{New ansatz2}) are valid.

The remaining equations of motion are Einstein's equations, which read 
in the unitary gauge
\begin{eqnarray}
&{}& \ddot{A} ( -2 \delta_\mu^0 \delta_\nu^0 - 2 \eta_{\mu\nu} )
+ (\dot{A})^2 ( 2 \delta_\mu^0 \delta_\nu^0 - \eta_{\mu\nu} )
+ \Lambda e^{2A} \eta_{\mu\nu} 
\nonumber\\
&=& \frac{1}{2} e^{2A} \eta_{\mu\nu} U 
- m^2 ( \delta_\mu^0 \delta_\nu^0 \frac{\partial U}{\partial X}
+ \delta_\mu^0 \delta_\nu^i \frac{\partial U}{\partial V^i}
+ \delta_\mu^i \delta_\nu^j \frac{\partial U}{\partial Y^{ij}} ).
\label{New Ein}
\end{eqnarray}
The requirement that these equations should have the same solutions 
as (\ref{generalized sol}) amounts to the similar conditions to
(\ref{Ansatz G}) and (\ref{GG}), which are
\begin{eqnarray}
U(\bar{X}, \bar{V^i}, \bar{Y}^{ij}) &=&  - K (m^2 e^{-2A})^n + 2 \Lambda,
\nonumber\\
\frac{\partial U(\bar{X}, \bar{V^i}, \bar{Y}^{ij})}{\partial Y^{ij}} 
&=&  - \frac{K}{3} n (m^2 e^{-2A})^{n-1} \delta_{ij}.
\label{New GG}
\end{eqnarray}
These conditions appear in a natural way when we assume the potential
term $U$ to be the polynomial type like (\ref{Ansatz G}), so we could
conclude that the classical solutions (\ref{generalized sol}) exist
as well in the most general massive gravity models by picking up an
appropriate form of the ghost potential.

\section{Analysis of propagating modes}

In this section, we analyse the physical modes propagating in the
cosmological backgrounds obtained in section 3. Thus, let us consider small
fluctuations around the metric 
\begin{eqnarray}
g_{\mu\nu} = e^{2A(t)} ( \eta_{\mu\nu} + h_{\mu\nu} ).
\label{Fluctuations}
\end{eqnarray}
With Eq. (\ref{Fluctuations}), some quantities relevant to the metric tensor
read in the second-order approximation level of the fluctuations $h_{\mu\nu}$
\begin{eqnarray}
g^{\mu\nu} &=& e^{-2A(t)} ( \eta^{\mu\nu}  - h^{\mu\nu} 
+ h^{\mu\alpha} h_\alpha^\nu),
\nonumber\\
\sqrt{-g} &=& e^{4A(t)} ( 1 + \frac{1}{2} h - \frac{1}{4} h_{\mu\nu} h^{\mu\nu}
+ \frac{1}{8} h^2 ),
\nonumber\\
\sqrt{-g} g^{\mu\nu} &=& e^{2A(t)} [ \eta^{\mu\nu} -  h^{\mu\nu}
+ \frac{1}{2} \eta^{\mu\nu} h + ( - \frac{1}{4} h_{\alpha\beta} h^{\alpha\beta}
+ \frac{1}{8} h^2 ) \eta^{\mu\nu} 
\nonumber\\
&-& \frac{1}{2} h h^{\mu\nu} + h^{\mu\alpha} h_\alpha^\nu ],
\label{Quantities}
\end{eqnarray}
etc. Here, summation over space-time indices $\mu = 0, 1, 2, 3$ is carried
out with the flat Minkowski metric $\eta_{\mu\nu}$ while that over
spatial indices $i = 1, 2, 3$ is done with the Kronecker $\delta_{ij}$,
so we have, for instance, 
\begin{eqnarray}
h = \eta_{\mu\nu} h^{\mu\nu} = - h_{00} + h_{ii}.
\label{h}
\end{eqnarray}

In the analysis of the physical modes, it is convenient to make use of 
the $(1+3)$-parametrization of the metric fluctuations \cite{Mukhanov}
\begin{eqnarray}
h_{00} &=& 2 \phi,
\nonumber\\
h_{0i} &=& S_i + \partial_i B,
\nonumber\\
h_{ij} &=& h_{ij}^{TT} - ( \partial_i F_j + \partial_j F_i )
- 2 ( \psi \delta_{ij} - \partial_i \partial_j E ),
\label{Parametrization}
\end{eqnarray}
where $h_{ij}^{TT}$ is a traceless, transverse spatial tensor
\begin{eqnarray}
\partial_i h_{ij}^{TT} = 0 = h_{ii}^{TT},
\label{tensor}
\end{eqnarray}
and $S_i$ and $F_i$ are transverse spatial vectors
\begin{eqnarray}
\partial_i S_i = 0 = \partial_i F_i.
\label{vector}
\end{eqnarray}
In this parametrization, the tensor $h_{ij}^{TT}$ has 2 degrees of freedom
(d.o.f.), the vectors $S_i$ and $F_i$ do $2 \times 2 = 4$ d.o.f., and
the scalars $\phi$, $B$, $\psi$ and $E$ do $1 \times 4 = 4$ d.o.f.,
so in total 10 d.o.f. which is exactly equal to the number of $h_{\mu\nu}$.

Incidentally, recall that in Einstein's general relativity only the massless 
spin 2 graviton $h_{ij}^{TT}$ of 2 d.o.f. is physical owing to general coordinate 
invariance. On the other hand, in a general massive gravity model \footnote{Here
the term $\it{general}$ means that there is no special residual gauge symmetry.}, 
the spin 2 tensor $h_{ij}^{TT}$ of 2 d.o.f., the spin 1 vector $F_i$ of 2 d.o.f. 
and the spin 1 scalar $\psi$ (or $E$) of 1 d.o.f. are physical so totally 
we have the massive graviton of 5 d.o.f. (the other modes are non-dynamical) 
which involves spins $\pm 2, \pm 1$ and $0$. 
The important point to notice is that there in general remains 
one scalar mode $E$ (or $\psi$), which has negative norm, so it is
in essence a $\it{ghost}$! This peculiar feature, that is, the existence
of a scalar ghost in the physical Hilbert space, is usually 
avoided by removing the ghost with the help of an enhancement of
gauge symmetry as in the Fierz-Pauli Lorentz-covariant massive gravity model 
\cite{Fierz}. However, this nice property is lost in curved backgrounds
owing to the disappearance of the enhanced gauge symmetry, and in
consequence we have the Boulware-Deser instability scalar mode \cite{Boulware}.

Accordingly, the real problem is to find a method of removal of the ghost mode 
from the physical spectrum without recourse to the enhanced gauge symmetry. 
In this article, we try to remove the ghost by appealing to the ghost 
condensation mechanism.

After a little lenghty calculation, the quadratic part of the Einstein-Hilbert
action and the cosmological term in (\ref{Main2}) is decomposed into tensor, 
vector and scalar sectors up to total derivative terms
\begin{eqnarray}
L_{EH}^{(2)} &=& L_{EH}^{(T)} + L_{EH}^{(V)} + L_{EH}^{(S)},
\nonumber\\
L_{\Lambda}^{(2)} &=& L_{\Lambda}^{(T)} + L_{\Lambda}^{(V)} + L_{\Lambda}^{(S)},
\label{EH+Lambda}
\end{eqnarray}
where each term is given by
\begin{eqnarray}
L_{EH}^{(T)} &=& e^{2A} [ - \{ \ddot{A} + \frac{1}{2} (\dot{A})^2 \}
(h_{ij}^{TT})^2 + \frac{1}{4} \{ (\partial_0 h_{ij}^{TT})^2
- (\partial_i h_{jk}^{TT})^2 \} ],
\nonumber\\
L_{EH}^{(V)} &=& e^{2A} [ 3 (\dot{A})^2 S_i^2 + 2 \{ \ddot{A} + \frac{1}{2} 
(\dot{A})^2 \} F_i \Delta F_i - \frac{1}{2} ( S_i + \partial_0 F_i)
\Delta ( S_i + \partial_0 F_i) ],
\nonumber\\
L_{EH}^{(S)} &=& e^{2A} [ (\dot{A})^2 \{ -9 \phi^2 - 9 \psi^2
+ 2 ( 9 \phi \psi - 3 \phi \Delta E + \frac{3}{2} (\partial_i B)^2 )
-2 ( \psi \Delta E + \frac{1}{2} (\Delta E)^2 ) \}
\nonumber\\
&+& \dot{A} ( 4 \phi \Delta B + 12 \phi \partial_0 \psi 
- 4 \phi \partial_0 \Delta E ) - 4 \ddot{A} ( \psi \Delta E + \frac{1}{2} (\Delta E)^2 )
\nonumber\\
&-& 4 \partial_0 \psi \Delta B - 4 \phi \Delta \psi + 6 \psi \partial_0^2 \psi
- 2 \psi \Delta \psi + 4 \partial_0 \psi \partial_0 \Delta E ],
\label{EH}
\end{eqnarray}
and
\begin{eqnarray}
L_{\Lambda}^{(T)} &=&  \frac{1}{2} \Lambda e^{4A} (h_{ij}^{TT})^2,
\nonumber\\
L_{\Lambda}^{(V)} &=& - \Lambda e^{4A} ( S_i^2 + F_i \Delta F_i ),
\nonumber\\
L_{\Lambda}^{(S)} &=& - 2 \Lambda e^{4A} [ - \frac{1}{2} \phi^2 
+ 3 \phi \psi - \phi \Delta E + \frac{3}{2} \psi^2
- \psi \Delta E - \frac{1}{2} (\Delta E)^2 + \frac{1}{2} (\partial_i B)^2 ],
\label{Lambda}
\end{eqnarray}
where $\Delta$ denotes the Laplacian operator defined as $\Delta \equiv
 \partial_i^2$.

Furthermore, provided that by $L_{F+G}$ we denote the sum of the quadratic part 
of the potential terms $L_F$ and $L_G$, it is also decomposed into each spin 
sector as
\begin{eqnarray}
L_{F+G}^{(T)} &=&  [ - \frac{1}{4} e^{4A} ( F - G ) 
- e^{2A} m^2 G_1 - m^4 G_2 ] (h_{ij}^{TT})^2,
\nonumber\\
L_{F+G}^{(V)} &=& [ \frac{1}{2} e^{4A} ( F - G ) 
+ 2 e^{2A} m^2 G_1 + 2 m^4 G_2 ] F_i \Delta F_i
+ \frac{1}{2} e^{4A} ( F - G ) S_i^2,
\nonumber\\
L_{F+G}^{(S)} &=&  e^{4A} [ - \frac{1}{2} ( F - G )
+ 2 F_{XX} \bar{X}^2 ] \phi^2 
+ \frac{1}{2} e^{4A} ( F - G ) (\partial_i B)^2
\nonumber\\
&+& [ e^{4A} ( F - G ) + 2 e^{2A} m^2 G_1 ] \phi (3 \psi 
- \Delta E) + [ \frac{3}{2} e^{4A} ( F - G ) 
+ 6 e^{2A} m^2 G_1 - 12 m^4 G_2 ] \psi^2
\nonumber\\
&+& [ - e^{4A} ( F - G ) - 4 e^{2A} m^2 G_1 + 8 m^4 G_2 ] \psi \Delta E
\nonumber\\
&+& [ - \frac{1}{2} e^{4A} ( F - G ) - 2 e^{2A} m^2 G_1 - 4 m^4 G_2 ]
(\Delta E)^2,
\label{F+G}  
\end{eqnarray}
where $F$, $F_{XX}$ and $G$ denote $F(\bar{X})$, $F_{XX}(\bar{X})$ and 
$G(\bar{W}^{ij})$, respectively.

We are now in a position to discuss each spin sector in order. 
In the tensor sector, the total Lagrangian takes the form
\begin{eqnarray}
L^{(T)} &\equiv& L_{EH}^{(T)} + L_{\Lambda}^{(T)} + L_{F+G}^{(T)}
\nonumber\\
&=& \frac{1}{4} e^{2A} [ (\partial_0 h_{ij}^{TT})^2
- (\partial_i h_{jk}^{TT})^2 ] 
+ [ - \frac{1}{2} e^{2A} m^2 G_1 - m^4 G_2 ] (h_{ij}^{TT})^2,
\label{tensor sector}
\end{eqnarray}
where the background equations, those are, Einstein's equations
(\ref{RE.Ein}) are used to simplify the expression. Then, the equations
of motion for $h_{ij}^{TT}$ give
\begin{eqnarray}
\Box h_{ij}^{TT} - 2 \dot{A} \partial_0 h_{ij}^{TT}
- 2 ( m^2 G_1 + 2 m^4 G_2 e^{-2A} ) h_{ij}^{TT} = 0,
\label{Eq for tensor}
\end{eqnarray}
where the d'Alembertian operator is defined as $\Box = \partial_\mu \partial^\mu
= - \partial_0^2 + \Delta$.

In order to see what mass of the graviton is, let us introduce
\begin{eqnarray}
\tilde h_{ij} = e^{A(t)} h_{ij}^{TT}.
\label{tilde h}
\end{eqnarray}
In terms of $\tilde h_{ij}$, the equations of motion  (\ref{Eq for tensor}) read
\begin{eqnarray}
\Box \tilde h_{ij} + [ \ddot{A} + (\dot{A})^2 - 2 ( m^2 G_1 + 2 m^4 G_2 e^{-2A}) ] 
\tilde h_{ij} = 0.
\label{Eq for tilde h}
\end{eqnarray}
Thus, the effective mass square of the tensor modes, $M_h^2$ is given by
\begin{eqnarray}
M_h^2 &=& - [ \ddot{A} + (\dot{A})^2 - 2 ( m^2 G_1 + 2 m^4 G_2 e^{-2A}) ] 
\nonumber\\
&=& \frac{4 n^2 + n - 2}{(n - 1)^2} \frac{1}{(t - t_0)^2},  
\label{Mass for h}
\end{eqnarray}
where we have used (\ref{generalized sol}) and (\ref{GG}) in the second
equality. 

Thus, the tensor modes in the cosmological backgrounds are
massive for $n > \frac{-1 + \sqrt{33}}{16}$ or $n < \frac{-1 - \sqrt{33}}{16}$
($n \ne 1$) while for $\frac{-1 - \sqrt{33}}{16} < n < \frac{-1 + \sqrt{33}}{16}$ 
they are effectively tachyonic and the backgrounds become unstable at least 
perturbatively. For $n = \frac{-1 \pm \sqrt{33}}{16}$, the graviton becomes
massless. Moreover, the effective mass of the graviton approaches zero
in the limit $t \rightarrow \infty$.

We next move to the vector sector. With the help of Einstein's
equations (\ref{RE.Ein}), the total Lagrangian for vector modes can be written as
\begin{eqnarray}
L^{(V)} &\equiv& L_{EH}^{(V)} + L_{\Lambda}^{(V)} + L_{F+G}^{(V)}
\nonumber\\
&=& - \frac{1}{2} e^{2A} ( S_i + \partial_0 F_i ) \Delta
( S_i + \partial_0 F_i ) + ( e^{2A} m^2 G_1 + 2 m^4 G_2 ) F_i 
\Delta F_i.
\label{vector sector}
\end{eqnarray}
Taking the variation with respect to $S_i$, we have the equation
\begin{eqnarray}
S_i = - \partial_0 F_i. 
\label{S-eq}
\end{eqnarray}
Substituting it into $L^{(V)}$, we have only the second term in (\ref{vector sector})
\begin{eqnarray}
L^{(V)} = ( e^{2A} m^2 G_1 + 2 m^4 G_2 ) F_i \Delta F_i.
\label{F-action}
\end{eqnarray}
The equations of motion for $F_i$ therefore yield
\begin{eqnarray}
F_i = 0.
\label{F-eq}
\end{eqnarray}
Plugging it back into $L^{(V)}$ in (\ref{F-action}) again, the Lagrangian
becomes identically vanishing, so that $F_i$ do not obey any equations
of motion and take any values. 

This arbitrariness, of course, is a consequence of the residual 
diffeomorphism invariance (\ref{Res-diff}). In this case,
the reparametrization symmetries read
\begin{eqnarray}
\delta h_{\mu\nu} = 2 \dot{A} \eta_{\mu\nu} \zeta^0 
+ \partial_\mu \zeta_\nu + \partial_\nu \zeta_\mu.
\label{Repra}
\end{eqnarray}
With $\zeta_i = \zeta_i(t)$ in (\ref{Res-diff}) and the parametrization
(\ref{Parametrization}), we have the residual gauge symmetries
\begin{eqnarray}
\delta S_i = \partial_0 \zeta_i, \
\delta F_i = - \zeta_i,
\label{Res-inv}
\end{eqnarray}
{}from which the modes $F_i$ and $S_i$ become non-dynamical.
More precisely speaking, it is $F_i$ that the residual gauge symmetries
(\ref{Res-inv}) make non-dynamical since $S_i$ remain non-dynamical 
irrespective of the existence of mass term as easily seen in (\ref{vector sector}).
In this way, we have proved that all the vector modes are not physical
but non-dynamical in this simple massive gravity model.

Finally, we are ready to examine the scalar sector, which is known to
be the most problematic and complicated. In fact, the Boulware-Deser
mode \cite{Boulware} appears in this sector in the Fierz-Pauli massive gravity model \cite{Fierz}.

After utilizing Einstein's equations (\ref{RE.Ein}), the total Lagrangian
is of form
\begin{eqnarray}
L^{(S)} &\equiv& L_{EH}^{(S)} + L_{\Lambda}^{(S)} + L_{F+G}^{(S)}
\nonumber\\
&=& 2 e^{2A} [ -2 \phi \Delta \psi - 2 \partial_0 \psi \Delta B
+ 2 \partial_0 \psi \partial_0 \Delta E + 3 \psi \partial_0^2 \psi
- \psi \Delta \psi
\nonumber\\
&+& \dot{A} ( 2 \phi \Delta B - 2 \phi \partial_0 \Delta E
+ 6 \phi \partial_0 \psi ) ]
+ e^{4A} ( F - G - 2 \Lambda + 2 F_{XX} \bar{X}^2 ) \phi^2 
\nonumber\\
&+& 3 [ e^{4A} ( F - G - 2 \Lambda ) + 2 e^{2A} m^2 G_1 - 4 m^4 G_2 ] \psi^2
+ 2 m^2 G_1 e^{2A} ( 3 \phi \psi - \phi \Delta E ) 
\nonumber\\
&-& 2 m^2 G_1 e^{2A} [ \psi \Delta E + \frac{1}{2} (\Delta E)^2 ]
+ 8 m^4 G_2 [ \psi \Delta E - \frac{1}{2} (\Delta E)^2 ].
\label{scalar sector}
\end{eqnarray}

The equation of motion for $B$ gives
\begin{eqnarray}
\partial_0 \psi = \dot{A} \phi.
\label{B}
\end{eqnarray}
Thus, integrating over $B$ and $\phi$ and using Einstein's equations again, 
up to total surface terms $L^{(S)}$ is reduced to 
\begin{eqnarray}
L^{(S)} 
&=& \frac{2}{(\dot{A})^2} m^4 F_{XX} \partial_0 \psi \partial_0 \psi
+ 2 n e^{2A} \psi \Delta \psi
- 3 n m^2 e^{2A} G_1 \psi^2
\nonumber\\
&-& \frac{2}{\dot{A}} e^{2A} m^2 G_1 \partial_0 \psi \Delta E
- 2 e^{2A} m^2 G_1 [ \psi \Delta E + \frac{1}{2} (\Delta E)^2 ]
\nonumber\\
&+& 8 m^4 G_2 [ \psi \Delta E - \frac{1}{2} (\Delta E)^2 ].
\label{scalar action}
\end{eqnarray}
Here the key point to understanding a scalar $\it{ghost}$ is that
there is no more $\partial_0 \psi \partial_0 \Delta E$ which is cancelled
when integrating over $B$ and $\phi$ (see Eqs. (\ref{B}) 
and (\ref{scalar sector})). Furthermore, the scalar mode
$E$ turns out to be non-dynamical since there is no time-derivative
term of $E$, so we can integrate over the mode $E$. Consequently,
$L^{(S)}$ becomes a Lagrangian for only a remaining scalar mode $\psi$.
A short calculation shows that
\begin{eqnarray}
L^{(S)} 
&=& \frac{2}{(\dot{A})^2} [ m^4 F_{XX} + \frac{n}{2n-1} \frac{K}{6}
m^{2n} e^{-2(n-2)A} ] \partial_0 \psi \partial_0 \psi
\nonumber\\
&+& 2 n e^{2A} \psi \Delta \psi - \frac{n^3}{2n-1} \frac{4 K}{3}
m^{2n} e^{-2(n-2)A} \psi^2.
\label{final scalar action}
\end{eqnarray}

This Lagrangian gives us several important information on $\psi$.
First, as far as ghost is concerned, the scalar mode $\psi$ never be a ghost 
when $I \equiv \frac{2}{(\dot{A})^2} [ m^4 F_{XX} + \frac{n}{2n-1} \frac{K}{6} 
m^{2n} e^{-2(n-2)A} ] > 0$.  Recall that $F_{XX} > 0$ from one of the ghost 
condensation ansatzes (\ref{Mini}), so $I$ is definitely positive as long as 
$n > \frac{1}{2}$. Moreover, owing to the overall factor $\frac{2}{(\dot{A})^2} 
= 2 (n-1)^2 (t - t_0)^2$, $I$ becomes divergent when $t \rightarrow \infty$
(we assume here that $F_{XX}$ is a positive constant as in the usual ghost condensation
model), implying that the mode $\psi$ is certainly dynamical.
Second, if we define the square of the mass of $\psi$ by
\begin{eqnarray}
M_\psi^2 = \frac{n^3}{2n-1} \frac{4 K}{3} m^{2n} e^{-2(n-1)A}, 
\label{Mass for psi}
\end{eqnarray}
it is also positive for $n > \frac{1}{2}$ and approaches zero in the limit
 $t \rightarrow \infty$ like the tensor modes. The region of $n > \frac{1}{2}$ 
is consistent with $n < 1$ for the positive acceleration and 
$n > \frac{-1 + \sqrt{33}}{16}$ for the positivity of the square of 
the graviton mass. 
Third, as seen easily in (\ref{final scalar action}), we have an unusal
time-dependent dispersion relation, so the scalar mode $\psi$ in general 
breaks the Lorentz invariance. Finally, in the case of $n = 0$, this model 
reduces to the original ghost condensation one.   

Hence, in this section, we have explicitly shown that only the propagating
modes around the cosmological backgrounds in this simple massive gravity
are the massive tensor modes $h_{ij}^{TT}$ and one scalar mode $\psi$.
As one peculiar feature, the massive scalar mode breaks the Lorentz
symmetry manifestly. Even if we have limited ourselves to the simple
massive gravity, we believe that this feature is also shared by the most
general massive gravity models.

\section{The absence of non-unitary mode in the 't Hooft model with ghost 
condensation}

In this section, for completeness, we shall present a proof of the absence
of the non-unitary mode in the 't Hooft model with ghost condensation
which was considered in section 2. 

Now let us start by considering the fluctuations around the unitary
gauge (\ref{Gauge}) for the four scalar fields
\begin{eqnarray}
\phi^a = m x^\mu \delta_\mu^a + \pi^a.
\label{Higgs}
\end{eqnarray}
Then, $X$ is expanded as
\begin{eqnarray}
X &=& - m^2 e^{-2A} + e^{-2A} ( - m^2 h_{00} + 2m \partial_0 \pi_0 )
\nonumber\\
&\equiv& \bar{X} + e^{-2A} ( - m^2 h_{00} + 2m \partial_0 \pi_0 ).
\label{X-fluct}
\end{eqnarray}

With Eq. (\ref{Higgs}), the $\phi^i$-equations of motion take the form
\begin{eqnarray}
\partial_\mu h^{\mu i} - \frac{1}{2} \partial^i h 
- \frac{1}{m} \Box \pi^i + 2 \dot{A} ( h^{0i} - \frac{1}{m}
\partial^0 \pi^i ) = 0.
\label{phi^i}
\end{eqnarray}
And the $\phi^0$-equation reads
\begin{eqnarray}
\partial_0 [ F_{XX}(\bar{X}) ( h_{00} - \frac{2}{m} 
\partial_0 \pi_0 ) ] = 0.
\label{phi^0}
\end{eqnarray}
At this stage, let us note that in the usual ghost condensation models
which satisfy the ansatzes (\ref{Mini}), one takes, for instance, 
the potential $F(X)$ to be $F(X) = c_1 ( X - \bar{X} )^2 + c_2$ with some 
constants $c_1>0$ and $c_2$. Then, it is natural to assume 
$\partial_0 F_{XX}(\bar{X}) = 0$. With this assumption, the $\phi^0$-equation 
reduces to
\begin{eqnarray}
\partial_0 ( h_{00} - \frac{2}{m} 
\partial_0 \pi_0 ) = 0.
\label{red-phi^0}
\end{eqnarray}
This equation is easily solved to 
\begin{eqnarray}
 h_{00} - \frac{2}{m} \partial_0 \pi_0 = C(x^i),
\label{sol-phi^0}
\end{eqnarray}
where $C(x^i)$ is an integration function depending on $x^i$ but
not $t$. We then find that $C(x^i)$ can be absorbed into the
definition of $\pi_0$ by redefining 
\begin{eqnarray}
\pi_0 \rightarrow \pi_0 + \frac{m}{2} t C(x^i).
\label{pi-redef}
\end{eqnarray}
Thus, we can take $C(x^i) = 0$. Then, $\pi_0$ can be expressed in
terms of $h_{00}$, so $\pi_0$ is not an independent degree of freedom
and can be neglected from the physical spectrum.

The remaining equations of motion which we have to examine are
Einstein's equations. At this point, for spatial diffeomorphisms,
we take the gauge conditions
\begin{eqnarray}
\pi^i = 0.
\label{pi-gauge}
\end{eqnarray}
With the help of the gauge conditions (\ref{pi-gauge}) and 
the $\phi^0$-equation (\ref{sol-phi^0}) with $C(x^i) = 0$,
Einstein's equations read
\begin{eqnarray}
&{}& \frac{1}{2} [ \partial_\rho \partial_\mu h_\nu^\rho 
+ \partial_\rho \partial_\nu h_\mu^\rho - \Box h_{\mu\nu}
- \partial_\mu \partial_\nu h - \eta_{\mu\nu} 
( \partial_\rho \partial_\sigma h^{\rho\sigma} - \Box h ) ]
\nonumber\\
&+& \frac{m^2}{2} [ 2 \delta_\mu^0 \delta_\nu^0 - ( 1 + h_{00} ) \eta_{\mu\nu}
- h_{\mu\nu} ]
\nonumber\\
&+& \dot{A} [ -2 \eta_{\mu\nu} ( \partial_\rho h^{0 \rho} 
+ \frac{1}{2} \partial_0 h ) + 2 \Gamma_{\mu\nu}^0(h) ]
\nonumber\\
&=& m^2 \delta_\mu^i \delta_\nu^j \delta_{ij} - \frac{m^2}{2} 
( 3 - h_{ii} ) \eta_{\mu\nu} - \frac{3}{2} m^2 h_{\mu\nu},
\label{ein-eq}
\end{eqnarray}
where $\Gamma_{\mu\nu}^0(h)$ denotes the Christoffel symbol in the
linear approximation of $h$.

In order to transform Einstein's equations (\ref{ein-eq}) to the more
tractable form, let us consider the following quantity
\begin{eqnarray}
P^\mu = \partial_\nu h^{\nu\mu} - \frac{1}{2} \partial^\mu h 
+ 2 \dot{A} h^{0\mu}.
\label{P}
\end{eqnarray}
With the $\phi^i$-equations (\ref{phi^i}) and the gauge conditions 
(\ref{pi-gauge}), we have 
\begin{eqnarray}
P^i = 0.
\label{zero P}
\end{eqnarray}
Under the general coordinate transformations (\ref{Repra}), the time-like
component of $P^\mu$ transforms as
\begin{eqnarray}
\delta P^0 = \Box \zeta_0 - 2 \dot{A} \partial_0 \zeta_0 - 2 m^2 \zeta_0.
\label{P^0}
\end{eqnarray}
Thus, using the remaining time-like diffeomorphism, we shall take a gauge
\footnote{The same gauge condition was also taken in \cite{Kaku2}.}
\begin{eqnarray}
P^0 = 0.
\label{zero P_0}
\end{eqnarray}
Then, a little thought or calculation shows that Einstein's equations
reduce to 
\begin{eqnarray}
\Box \hat{h}_{\mu\nu} - 2 \dot{A} \partial_0 \hat{h}_{\mu\nu}
=  2m^2 \hat{h}_{\mu\nu},
\label{ein-eq2}
\end{eqnarray}
where we have defined $\hat{h}_{\mu\nu} \equiv h_{\mu\nu}
- \frac{1}{2} \eta_{\mu\nu} h$. 

Notice that the right-hand side of $\delta P^0$ in Eq. (\ref{P^0})
has the same expression as the equations of motion for $\hat{h}_{\mu\nu}$.
Given that $\Box \zeta_0 - 2 \dot{A} \partial_0 \zeta_0 - 2 m^2 \zeta_0
= 0$, we obtain $\delta P^0 = 0$. In other words, there remains a
residual gauge symmetry associated with such the $\zeta_0$.
The existence of the residual symmetry makes it possible to take
a gauge
\begin{eqnarray}
h = 0,
\label{h-gauge}
\end{eqnarray}
{}from which we have the equation 
\begin{eqnarray}
\hat{h}_{\mu\nu} = h_{\mu\nu}.
\label{hat h}
\end{eqnarray}

Finally, introducing $H_{\mu\nu} \equiv e^{A(t)} h_{\mu\nu}$, it turns out 
that the whole equations read
\begin{eqnarray}
\Box H_{\mu\nu} - \frac{3}{2} m^2 H_{\mu\nu} &=& 0,
\nonumber\\
\partial^\nu H_{\mu\nu} - \dot{A} H_{0\mu} &=& 0,
\nonumber\\
H &=& 0,
\nonumber\\
e^{-A} H_{00} - \frac{2}{m} \partial_0 \pi_0 &=& 0.
\label{whole eq}
\end{eqnarray}
These equations show that the graviton has mass $\sqrt{\frac{3}{2}} m$
and the same 5 degrees of freedom as usual massive graviton modes.
Moreover, there is no non-unitary mode since $\pi_0$ mode
is expressed in terms of $H_{00}$. In this way, we find that
the 't Hooft model with ghost condensation describes a physically
plausible massive gravity model in the linearly expanding universe with zero
acceleration.

\section{Conclusions and Discussion}

In this paper, we have shown that the 't Hooft model with ghost condensation
is free of non-unitary scalar mode and is a massive gravity model in
the linearly expanding cosmological universe with zero acceleration.
This proof is rather general and simple. The reason is that in this model the
acceleration is vanishing, from which many of equations take tractable
expressions. Notice that the situation where there is no acceleration
in the Friedmann-Robertson-Walker metric is similar to that where the
equation of state is $P = - \frac{1}{3} \rho$ with $P$ and $\rho$ being
respectively the pressure and the matter density. This analogy might be
useful for a better understanding of this solution.   

Furthermore, we have showed that a more general massive gravity
model has an interesting class
of classical solutions with the property of polynomially expanding cosmological 
universes with non-zero acceleration. This class of solutions is classified by
a constant $n$ in the potential. Recall that this constant $n$ is not a completely
free parameter but receives some restriction from physical conditions. The first
requirement that the model should describe a positive acceleration leads to 
$n < 1$. The second requirement that the massive graviton should not be a tachyon 
gives us a condition $n > \frac{-1 + \sqrt{33}}{16}$ or $n < \frac{-1 - \sqrt{33}}{16}$
($n \ne 1$). Moreover, the third requirement that the scalar mode should not be
a ghost and/or a tachyon provides a final condition $n > \frac{1}{2}$.
As a result, there exists the parameter region $\frac{1}{2} < n < 1$,
which satisfies three requirements above at the same time.

The existence of non-zero acceleration in the polynomially expanding cosmological 
universes has made it difficult to prove that this massive gravity is free of the non-unitary 
mode and the Boulware-Deser instability. In order to clarify the physical propagating modes, 
we have used the (1+3)-parametrization of the metric fluctuations. Using this parametrization,
it has been explicitly shown that the tensor modes become massive, 
the vector modes are non-dynamical, three of the scalar modes are also non-dynamical
and one scalar becomes massive. This scalar is originally non-unitary mode,
that is, a ghost, but becomes a normal particle because of the ghost
condensation mechanism. However, the dispersion relation is not usual
one, so this mode breaks the Lorentz invariance as expected from
the ghost condensation scenario.

In most of models which attempt to explain the late-time cosmic acceleration,
the acceleration is driven by some exotic matter with negative pressure
called dark matter. On the other hand, in the models considered in this
paper, the acceleration is driven by the massive graviton and an extra scalar
which are originally part of components of the metric tensor $h_{\mu\nu}$.
This fact is in sharp contrast to the models which have been proposed 
so far. To our knowledge, no serious attempts have been made to study
the late-time cosmic acceleration this way. 

As future's problems, we first wish to construct a Lorentz-invariant massive gravity model
in a flat Minkowski space-time such that the model is free of the non-unitary mode 
since such a model describes a world of QCD. We also wish to examine various
phenomenological aspects of the models that we have considered in this paper.
It is known that the models lead to interesting phenomenology around a flat Minkowski
background \cite{Dubovsky1}-\cite{Rubakov2}, so we think that the models around our cosmological 
backgrounds also give us new definite predictions for a scenario for inflation 
and density perturbations.

\vs 1   

\end{document}